# A COMPLEX DATA WAREHOUSE FOR PERSONALIZED, ANTICIPATIVE MEDICINE

Jérôme Darmont and Emerson Olivier

ERIC, University of Lyon 2
5 avenue Pierre Mendès-France
69676 Bron Cedex, FRANCE

Phone: +33 478 774 403, Fax: +33 478 772 378, E-mail: jerome.darmont@univ-lyon2.fr

## INTRODUCTION

With the growing use of new technologies, healthcare is nowadays undergoing significant changes. The development of electronic health records shall indeed help enforcing personalized, lifetime healthcare and pre-symptomatic treatment, as well as various analyses over a given population of patients. Such an information-based medicine has to exploit medical decision-support systems and requires the analysis of various, heterogeneous data, such as patient records, medical images, biological analysis results, etc. (Saad, 2004).

In this context, we work with a physician on large amounts of data relating to high-level athletes. His aim is to make the analyzed subjects the managers of their own health capital, by issuing recommendations regarding, e.g., life style, nutrition, or physical activity. This is meant to support personalized, anticipative medicine. To achieve this goal, a decision-support system must allow transverse analyses of a given population and the storage of global medical data such as biometrical, biological, cardio-vascular, clinical, and psychological data.

Data warehousing technologies are now considered mature and can form the base of such a decision-support system. Though they primarily allow the analysis of numerical data, the concepts of data warehousing remain valid for what we term *complex data*. In this context, the warehouse measures, though not necessarily numerical, remain the indicators for analysis, and analysis is still performed following different perspectives represented by dimensions. Large data volumes and their dating are other arguments in favor of this approach. Data warehousing can also support various types of analysis, such as statistical reporting, on-line analysis (OLAP) and data mining.

In this paper, we present the design of the complex data warehouse relating to high-level athletes. It is original in two ways. First, it is aimed at storing complex medical data. Second, it is designed to allow innovative and quite different kinds of analyses to support:
1. personalized and anticipative medicine (in opposition to curative medicine) for well-identified patients;
2. broad-band statistical studies over a given population of patients. Furthermore, the system includes data relating to several medical fields.

It is also designed to be evolutionary to take into account future advances in medical research.

## WAREHOUSE MODEL

### Global Architecture

To make our solution evolutionary, we adopted a bus architecture (Kimball and Ross, 2002). It is composed of a set of conformed dimensions and standardized definitions of facts. In this framework, the warehoused data related to every medical field we need to take into account represent datamarts that are plugged into the data warehouse bus and receive the dimension and fact tables they need. The union of these datamarts may be viewed as the whole data warehouse.

Figure 1 represents the global architecture of our data warehouse. Straight squares symbolize fact tables, round squares symbolize dimensions, dotted lines embed the different datamarts, and the data warehouse bus is a rounded rectangle. It is constituted by dimensions that are common to several datamarts. The main dimensions that are common to all our datamarts are patient, data provider, time, and medical analysis (that regroups several kinds of analyses). Of course, some datamarts (such as the cardio-vascular datamart) do have specific dimensions that are not shared. Our data warehouse also includes a medical background datamart (not depicted), and more datamarts are currently being developed.

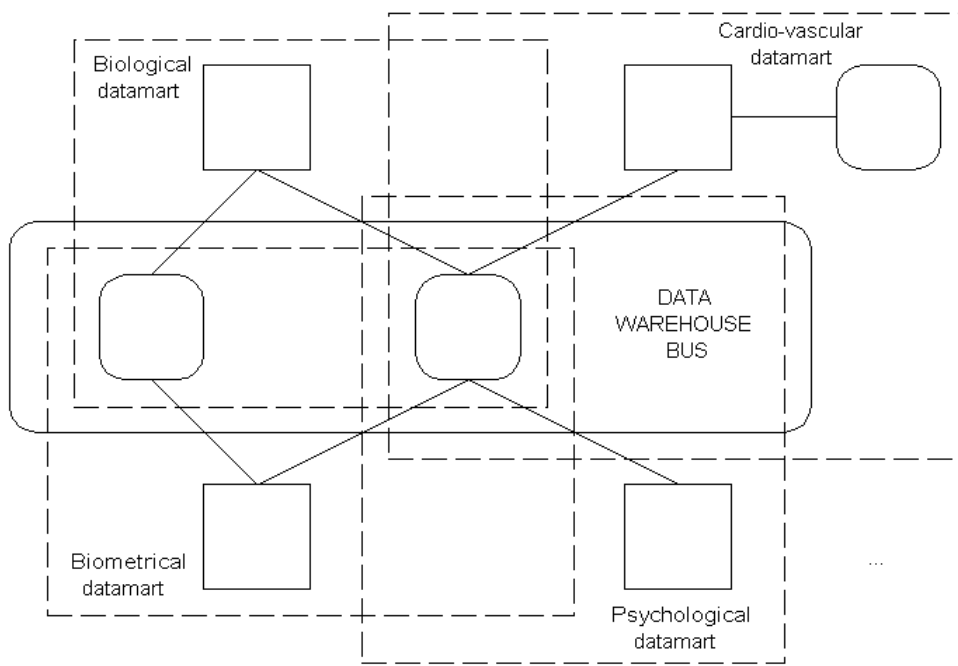

Figure 1: Data warehouse global architecture

Simple Datamarts: the Biological and Biometrical Datamarts

Input biological data are the results of various biological examinations (biochemistry, protein checkup, hematology, immunology, toxicology…), which are themselves subdivided into actual analyses (e.g., an hematology examination consists in reticulocyte numbering and an hemogram). These data are available under the form of unnormalized spreadsheet files from different sources. They are thus heterogeneous and often refer to the same examination using different terms or abbreviations, use different units for the same numerical values, etc. This heterogeneity is dealt with during the ETL (Extract, Transform, Load) process.

Biometrical data are measured during medical examinations. They include weight, height, pulse rate, fat percentage, and blood pressure. Though their structure is simpler than that of the biological data, they require a fine granularity. For example, the weight of an athlete may be measured twice a day, before and after training. This has an impact on data warehouse modeling. More precisely, it helps defining the granularity of the time dimension hierarchy (see below).

Figure 2 represents the architecture of our biological and biometrical datamarts. The biological fact table stores an exam result under the form of a numerical value (e.g., a reticulocyte numbering). It is described by four dimensions: patient, time of the examination, data provider (typically the laboratory performing the analysis), and the analysis itself. The biometrical fact table stores numerical biometrical values (e.g., weight). It is described by the same dimensions than the biological datamart, the "analysis" actually representing a measurement. The patient, data provider, time, and medical examination dimensions are thus all shared, conformed dimensions. Attributes in dimension tables are not detailed due to confidentiality constraints.

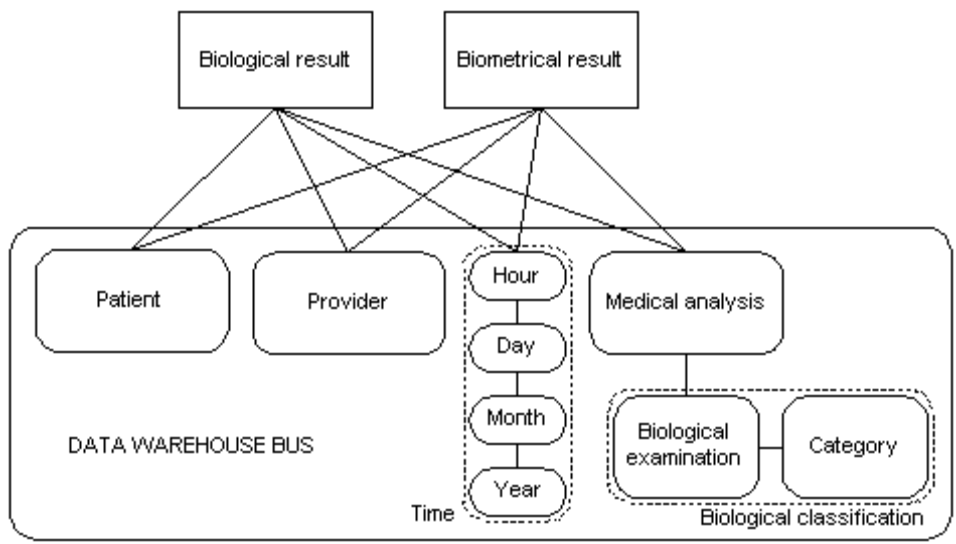

Figure 2: Biological and biometrical datamarts' architecture

Note that the medical analysis and time dimensions are further organized into hierarchies, to take into account the particularities identified into the source data. Here, biological and biometrical data are distinguished: the simple biometrical data are not normalized nor organized in a hierarchy, while the biological data are. Hence, the description of biometrical facts only appear in the medical analysis dimension table, while biological facts are further described by a hierarchy of biological examinations and categories.

Each datamart is thus modeled as a snowflake schema rather than as a simpler, classical star schema. Since the biological and biometrical fact tables share their dimensions, our overall data warehouse follows a constellation schema. In such an architecture, it is easy to add in new fact tables described by existing dimensions.

Finally, our data warehouse also includes metadata that help managing both the integration of source data into the warehouse (e.g., correspondence between different labels or numerical units, the French SLBC biomedical nomenclature, etc.) and their exploitation during analysis (e.g., the interval between which an examination result is considered normal).

Complex Datamart: the Cardio-Vascular Datamart

Figure 3 represents the architecture of our cardio-vascular datamart. Here, the complex nature of source data, which are constituted of raw measurements (e.g., ventricle size), multimedia documents (e.g., echocardiograms) and a conclusion by a physician, cannot be embedded in a single, standard fact table. Hence, we exploit a set of interrelated tables that together represent the facts. They are represented as dotted, straight squares.

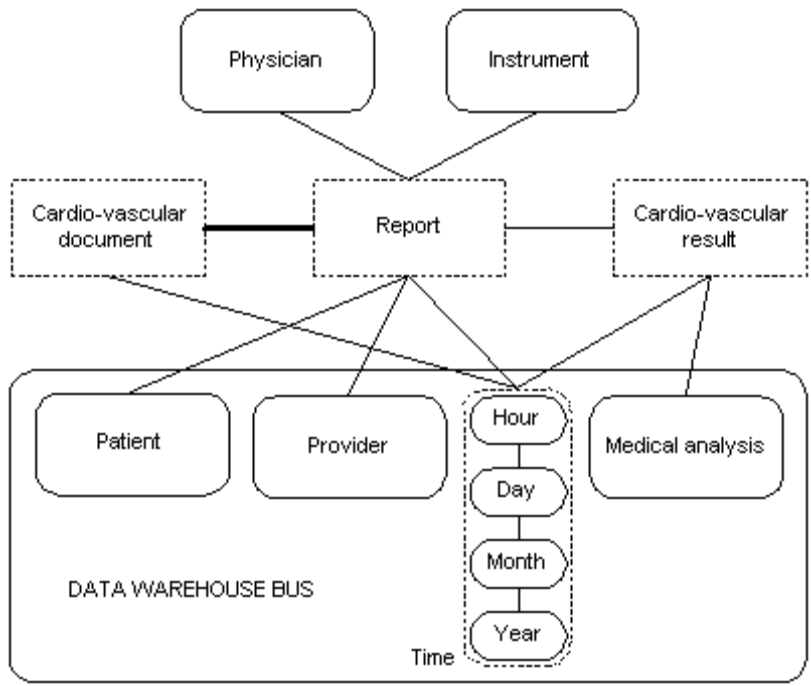

Figure 3: Cardio-vascular datamart's architecture

The report mainly contains the physician's conclusion. It is the central element in our "complex fact". It is linked to several numerical analysis results that help building the physician's conclusion. It is also related to multimedia documents such as medical images that also help devising the diagnosis. Note that this relationship, which is represented as a bold line, is a many-to-many relationship. Some documents may indeed be referred to by several reports, for instance to take into account a patient's evolution over time, through distinct series of echocardiograms. Each component of our "complex fact" may be individually linked to the dimensions. Cardio-vascular documents and results are indeed not descriptors of the would-be fact: report, they are part of a fuzzier fact that is composed of several entities.

Finally, note that cardio-vascular documents cannot currently be exploited by OLAP technologies. However, we need to store them and have them accessible for medico-legal reasons. Furthermore, extensions to the usual OLAP operators should make this exploitation possible in the near future (Ben Messaoud *et al.*, 2004).

RELATED WORK

The first family of medical data warehouses we identify are repositories tailored for later data mining treatments. Since data mining techniques take "attribute-value" tables as input, some of these warehouses do not bear a multidimensional architecture (Prather *et al.*, 1997; Sun *et al.*, 2004), and might prove less evolutionary than the solution we selected.

A second family is constituted of biological data warehouses that focus on molecular biology and genetics (Schönbach *et al.*, 2000; Eriksson and Tsuritani 2003; Sun *et al.*, 2004), which bear interesting characteristics. For instance, some of them include metadata and ontologies from various public sources such as RefSeq or Medline (Engström and Asthorsso, 2003; Shah *et al.*, 2005). The incremental maintenance and evolution of the warehouse is also addressed (Engström and Asthorsso, 2003). However, the particular focus of these approaches makes them inappropriate to our current needs, which are both different and much more diversified. The future developments of our data warehouse will probably exploit this existing work, though.

To sum up, the existing medical warehouse that is closest to our own is a cardiology data warehouse (Tchounikine *et al.*, 2001; Miquel and Tchounikine 2002). Its aim is to ease medical data mining by integrating data and processes into a single warehouse. However, raw sensor data (e.g., electrocardiograms) are stored separately from multidimensional data (e.g., patient identity, therapeutic data), while we seek to integrate them all in our cardio-vascular datamart.

RESEARCH PERSPECTIVES

This work opens up two kinds of perspectives. The first one concerns the actual contents and significance of the data warehouse. It involves modeling and adding in new datamarts. The cardio-vascular and psychology datamarts are already implemented and more are currently being developed (such as the medical background datamart) or scheduled. This helps broadening the scopes of analyses. Other output than statistical reports are also envisaged. Since we adopted a dimensional modeling, OLAP navigation is also definitely possible, and "attribute-value" views could easily be extracted from the data warehouse to allow data mining explorations.

The second kind of perspectives is more technical and aim at improving our prototype. This includes automating and generalizing the ETL process on all the datamarts, which is currently an ongoing task. We also follow other leads to improve the user-friendliness of our interfaces and the security of the whole system, which is particularly primordial when dealing with medical, personal data.

ACKNOWLEDGMENTS

The authors thank Dr Jean-Marcel Ferret, the promoter of the personalized and anticipative medicine project. This work has been co-funded by the Rhône-Alpes Region.